\documentstyle[12pt,epsf]{article}
\newcommand{\be}{\begin{equation}}
\newcommand{\ee}{\end{equation}}
\newcommand{\bea}{\begin{eqnarray}}
\newcommand{\eea}{\end{eqnarray}}
\newcommand{\ci}{\cite}
\newcommand{\bi}{\bibitem}
\newcommand{\nono}{\nonumber \\}

\newcommand{\dd}{\partial}

\newcommand{\half}{\frac{1}{2}}

\def\dal{\,\lower0.3ex\vbox{\hrule\hbox{\vrule\kern2pt\vbox{\kern4pt\kern4pt}
\kern2pt\vrule}\hrule}\,}

\def\o{\omega}
\def\L{{\cal L}}

\begin{document}
\title{{\bf Trapped states and bound states of a soliton in a well}}
\vspace{1 true cm}
\author{G. K\"albermann\thanks{Permanent address: Faculty of
Agriculture and Racah Institute of Physics, Hebrew University, Jerusalem 91904
, Israel. E-mail address: Kalbermann@comp.tamu.edu}\\
Cyclotron Institute\\
Texas A\&M University\\
College Station, TX 77843, USA\\}

\maketitle

\begin{abstract}
\baselineskip 1.5 pc
The nature of the interaction of a soliton with an attractive 
well is elucidated using
a model of two interacting point particles.
The model explains the existence of trapped states
at positive kinetic energy, as well as reflection by
an attractive impurity.
The transition from a trapped soliton state to a bound state is studied.
Bound states of the soliton in a well are also found.
\end{abstract}

\newpage
\baselineskip 1.5 pc

Topological solitons arise as nontrivial solutions
in field theories with nonlinear interactions.
These solutions are stable against dispersion. 
Topology enters through the absolute conservation of a topological
charge, or winding number.

It is for this reason they become so important in the description
of phenomena like, optical self-focusing,
magnetic flux in Josephson junctions\ci{shen}
or even the very existence of stable elementary
particles, such as the skyrmion \ci{sk,an}, as a model of hadrons.

Interactions of solitons with external agents become
extremely important. These interactions
allow us to test the validity of such models in real situations.

In a previous work \ci{kal97} the interaction of a 
soliton in one space dimension with 
finite size impurities was investigated. 

In the works of Kivshar et al.\ci{kivshar} (see also ref.~\ci{cuba,cuba1}),
it was found that the soliton displays
unique phenomena when it interacts with an external impurity.
The existence of trapped solutions
for positive energy or, bound states in the continuum, is
a very distinctive effect for the soliton in interaction
with an attractive well.

We can understand the origin of impurity interactions
of a soliton by looking at the impurity as a nontrivial
medium in which the soliton propagates.
An easy way to visualize these interactions consists in 
introducing a nontrivial metric
for the relevant spacetime. 
The metric carries the
information of the medium characteristics.

Consider for example a 1+1 dimensional scalar field theory 
supporting topological solitons in flat space, 
immersed in a backgound determined by the metric $g_{\mu\nu}$.
The standard manner of coupling the scalar
field to the metric is
\be\label{lag}
\L  = \sqrt{g}\bigg[g^{\mu\nu}\half \dd_\mu\phi\,\dd_\nu\phi
- U(\phi)\big]
\ee
where $g$ is the of the determinant of the metric, and
U is the self-interaction that enables the existence of
the soliton. For a weak potential we have\ci{rob}

\bea\label{metric}
{g_{00}}&\approx&1+V(x)\nono
{g_{11}}&=&-1\nono
{g_{-11}}&=&g_{1-1}=0
\eea

Where $V(x)$ is the external space dependent potential.
The equation of motion of the soliton in the background space becomes
\be\label{eq1}
\frac{{\dd}^2 \phi}{\dd t^2} -{\sqrt{g}}^{-1}
 \frac{\dd}{\dd x}\big[\sqrt{g}\frac{\dd\phi}{\dd x}\bigg]\nono 
+ g_{00}\frac{\dd U}{\dd\phi}=0.
\ee
This equation is identical, for slowly varying
potentials, to the equation of motion of a soliton interacting with
an impurity $V(x)$. 
Impurity interactions are therefore acceptable couplings of
a soliton to an external potential.
It is also the only way to couple the soliton without spoiling
the topological boundary conditions.

The interaction of a soliton with an attractive impurity
shows, however, some puzzling effects\ci{kivshar,kal97}.
A soliton can be trapped in it, when it impinges onto the well
with positive kinetic energy. 
Energy conservation demands that
the soliton fluctuates and distorts in trapped states inside the well.
Even more counterintuitive is the fact that the soliton can be reflected
by the well.

Neither of these effects are possible for classical
point particles.
The difference must obviously be due to the extended
character of the soliton. 
We should then be able to reproduce these effects with a 
classical model for an extended object.

The simplest extended object one
can envisage consists of two classical point particles connected by
a massless spring. A repulsive force between them is also needed to prevent
their collapse to zero size.

Consider such a system where each particle interacts also
with an external attractive well.
The classical nonrelativistic one-dimensional lagrangian for
the system of equal masses $m_1=m_2=1$ becomes:
\be\label{clas1}
{\L}_{sys}  = \frac{\dot{x}_1^2}{2}+\frac{\dot{x}_2^2}{2}-
k~\frac{(x_1-x_2)^2}{2}-\frac{\alpha}{{|x_1-x_2|}^n}
+V(x_1)+V(x_2)
\ee
For the potential well we take
\be\label{well}
V(x)~=A~e^{-\beta~x^2}
\ee
We prepare the two-particle system at rest at a large distance
far away from the well with  an initial speed $v$. 
The equilibrium interparticle separation is ${x_0}^{n+2}~=~\frac
{n~\alpha}{k}$.

The equations of motion are not solvable analytically.
Using the numerical algorithm used in ref.~{\ci{kal97}}
we can find the outcome of the scattering events as a 
function of the initial speed.
\begin{figure}[tb]
\epsffile{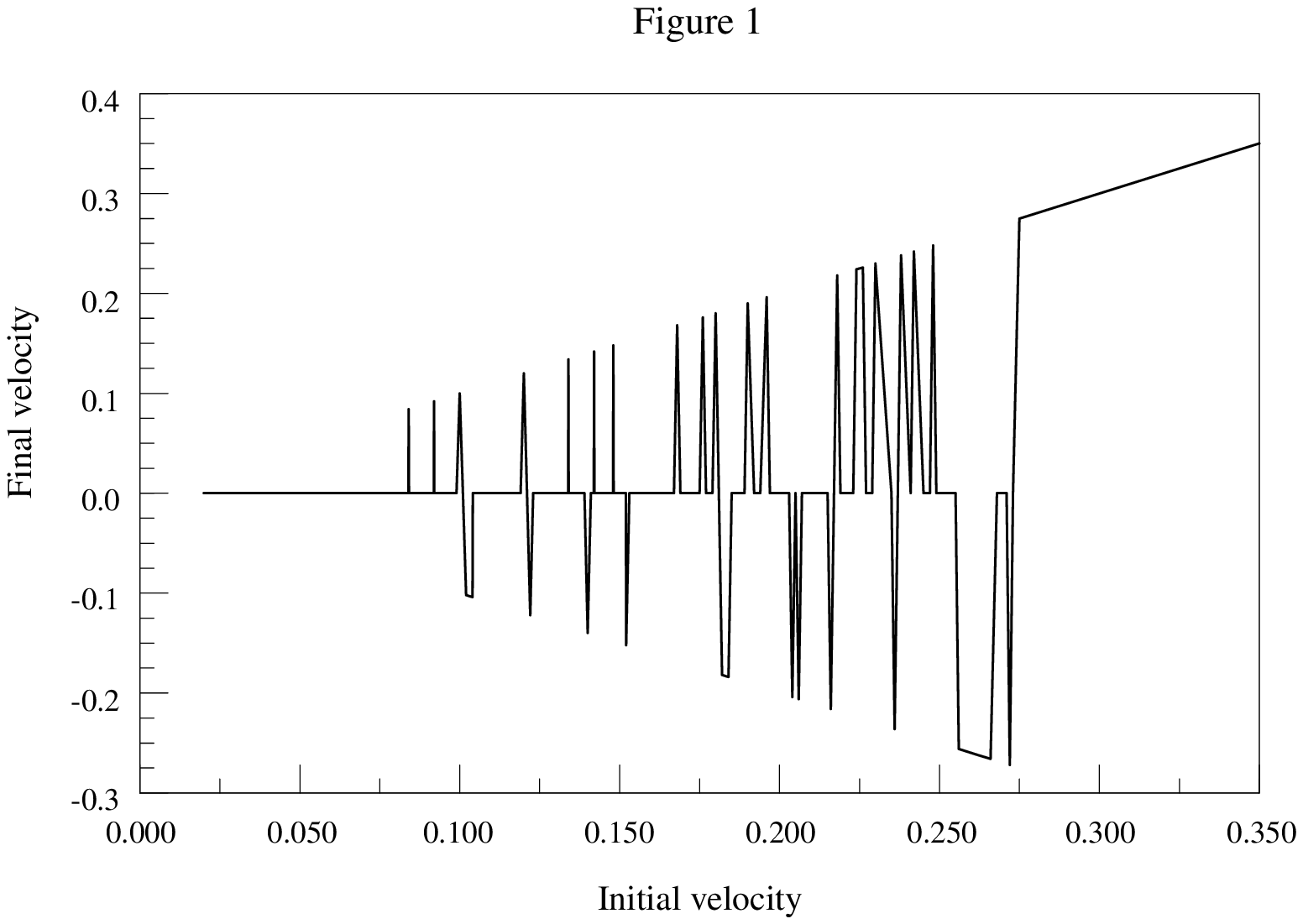}
\vsize=5 cm
\label{fig1}
\end{figure}

Figure 1 exemplifies the results
for the choice of parameters $k=1,~\alpha=1,~n=2,~A=2,~\beta=1$.

Quite unexpectedly, it is found that the system behaves exactly
like the soliton.

The system can be trapped, reflected or transmitted through the well
by changing the initial speed. 

When the system is trapped, it oscillates with a null 
average speed, the kinetic energy stored in the vibrational and
deformation modes.

Minute changes of the initial speed around a value leading to a trapped state,
may generate reflection or transmission events.

The effects are independent of
the functional dependence of the interactions and external potential,
and of the values of the parameters. 
It looks as if the behavior is universal.

In figure 1 we used a grid for $v$ of $dv=.001$. 
Using a finer
grid, each region of reflection-transmission unfolds to more islands
of trapping, reflection and transmission.

Finer and finer grids show more and more structure. 

Figure 2 shows a detailed expansion of the velocity range around v=.12 with 
a grid dv=.0002. 
The system is chaotic, an infinitesimal change in the initial speed
produces diverging results.

Many of the phenomena related to chaotic behavior may
be identified in the system. Scaling and bifurcation are evident
and perhaps even fractal structure.
(This issue will be taken up in another work)

\begin{figure}[tb]
\epsffile{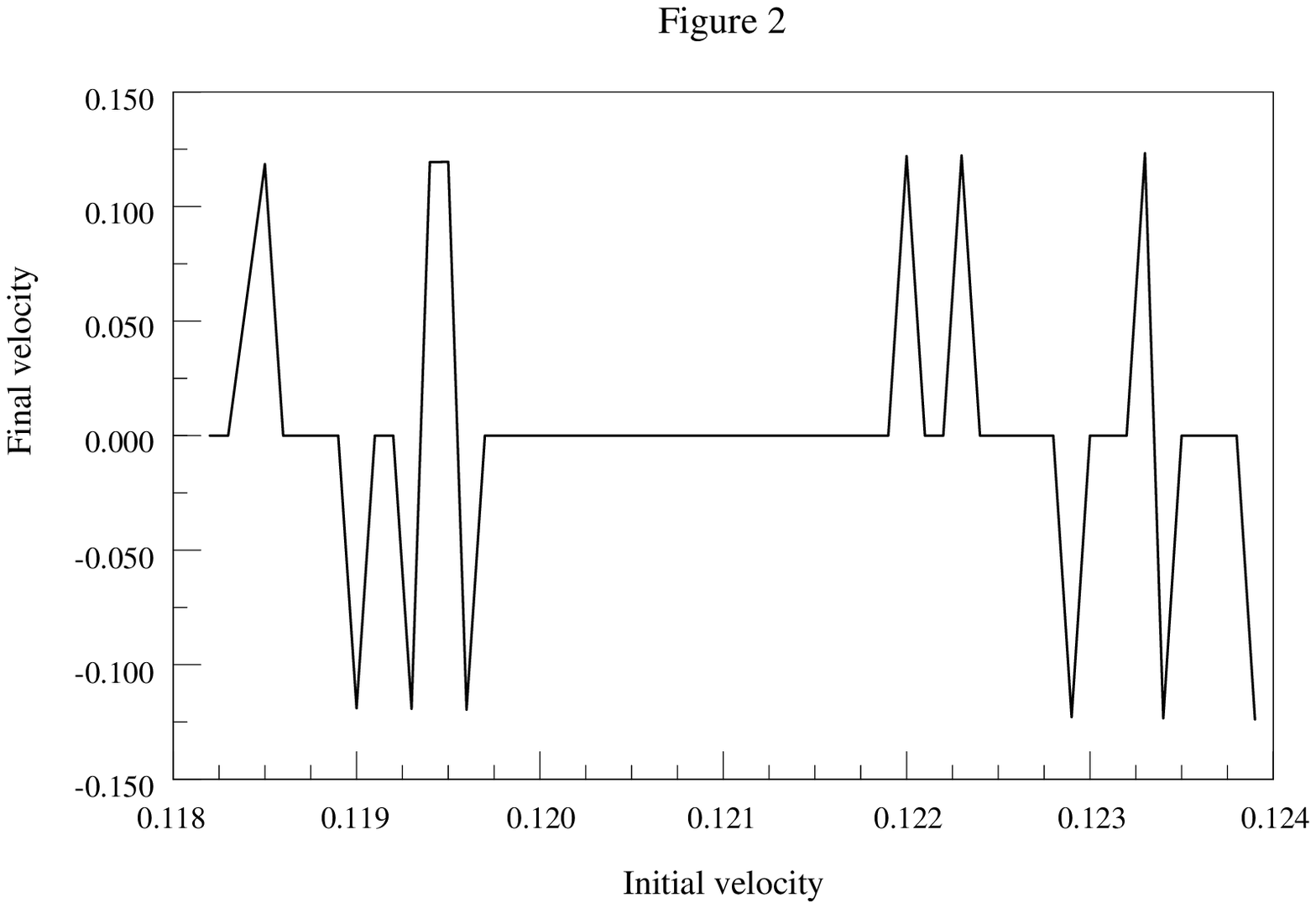}
\label{fig2}
\end{figure}

It is now safe to claim that the unexpected behavior of a soliton interacting 
with an attractive well may be traced back to its extended nature.
If we consider each ${\phi}(x)$ as a classical pointlike
object we will find interactions between neighboring particles
of attractive and repulsive character. The basic attractive interaction is
provided by the space derivative of the soliton lagrangian and 
a piece of the self-interaction potential, while the repulsive 
interaction is provided by the latter only.
\newpage
We now focus on the fate of a trapped soliton state.

Consider the kink lagrangian
\be\label{lag1}
 {\cal L}  = \half \dd_{\mu}\phi\,\dd^{\mu}\phi-
{1 \over 4 }\Lambda~{\bigg(\phi^2 - {m^2\over \lambda}\bigg)}^{2}
\ee
Here 
\be \label{Lambda}
\Lambda = \lambda + V(x)
\ee
$\lambda$ being a constant, and $V(x)$ the impurity potential\ci{kal97}
\be \label{u}
V(x) = h~{cosh^{-2}\bigg({a~(x-x_c)}\bigg)}
\ee

Independently of the choice of parameters it is found that
trapped states decay. 
The soliton radiates energy and consequently the amplitude
of the oscillations decreases.
The trapped states become asymptotically bound states.

Figure 3 shows the amplitude of the oscillation of the soliton,
namely, the value of the field at the center of the well,
as a function of time. Here we used
$m=1,~\lambda=1,~h=-3,~a=2,~x_1=3~$. 
The soliton impinges from the left. The initial location of the 
center of the soliton is chosen to be far enough 
from the well at $x=-3$, with an initial speed
$v=.025$.

In order to visualize the decay and emission of
radiation we extended the x-axis to $-140~\leq x~\leq 140$ with 
a grid of dx= 0.1. This coordinate span allows for
radiation to progate for a long distance away from the
trapping zone without being reflected.
\begin{figure}[tb]
\epsffile{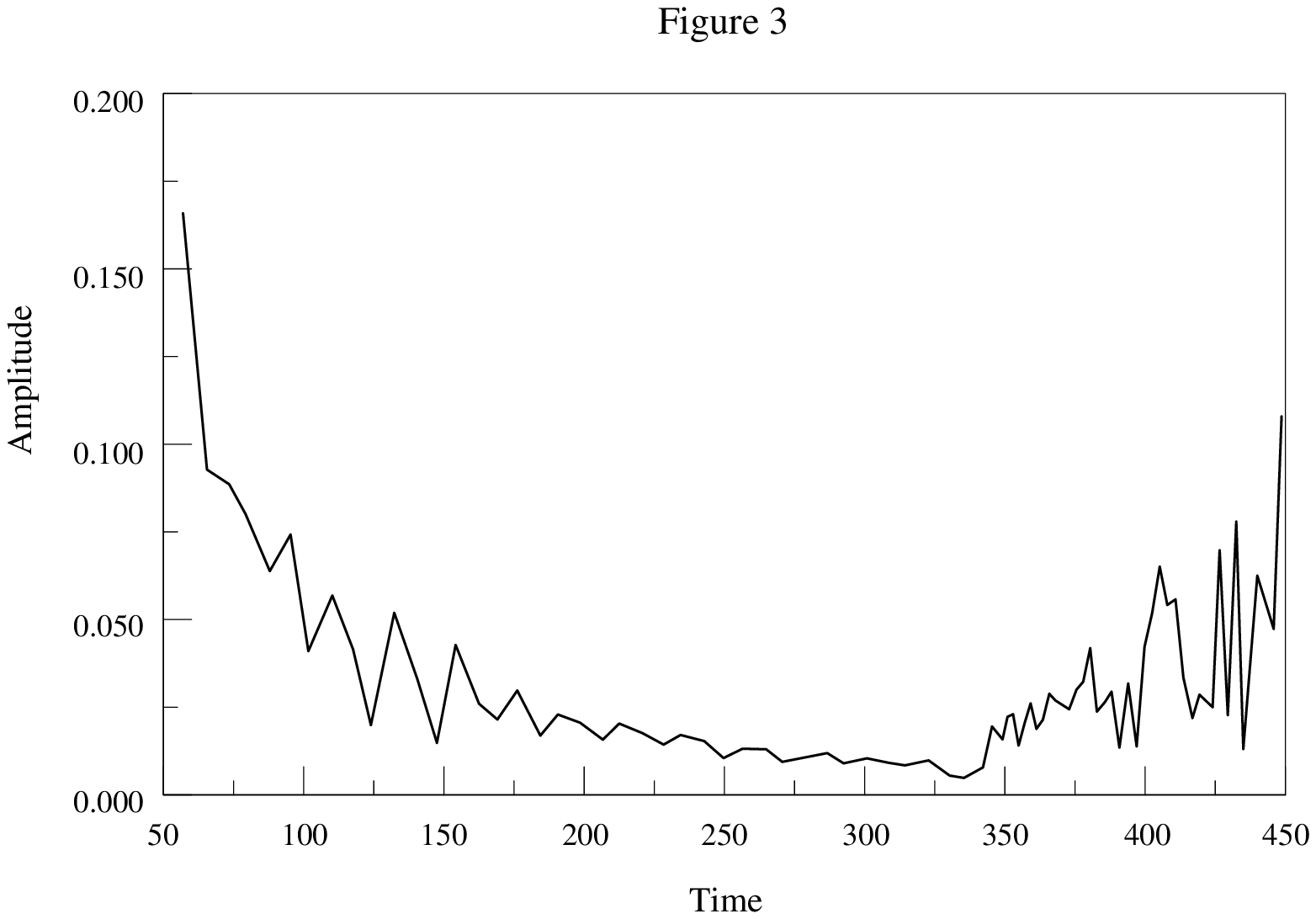}
\label{fig3}
\end{figure}

When the soliton reaches the well, it oscillates
and starts to emit radiation. The emission of radiation
damps the oscillations.
After a certain time, and due to
the finite extent of the x-axis, radiation reflects back
from the boundaries and reaches the soliton. The soliton
subsequently absorbs the radiation and its amplitude starts to
increase.
The time taken for radiation to return to the soliton
is the travel time for the fastest 'mesons' of the theory.

The dispersion relation for the radiated mesons can be extracted
from the expansion of the scalar field around the soliton solution. 
Using $\lambda~=~m~=1$ we find $\o^2=k^2+2$. 
The velocity of
the mesons is bounded by 

$u_{max}=\frac{\o}{k}\bigg)_{max} = 1$. This
is clearly observed in figure 3. The reabsorption of radiation starts
after the first mesons arrive back from the boundaries
to the well.
The distance between the well and the boundary
is 140, therefore $t_{absorption}=280/u_{max}=280$

The frequency of the oscillations
of the soliton in a trapped state may be estimated analytically.
Using an expansion of the potential in eq.~(\ref{u})
around the bottom of the well
$V(x)\approx-V_0+\epsilon~y^2,~ y=x-x_c$ and an ansatz appropriate for
small oscillations of the soliton around the center of the well
$\phi\approx(y+\delta~y^3/2)~sin(\o(t-t_0))$ we find $\o^2=2~\mu$.

With $\mu\approx\pm\sqrt{~\frac{4}{5}\epsilon} + \frac{9}{10}(V_0-1)$.
(The positive solution has to be chosen)

The formula compares reasonably well with the leading frequency
of oscillation of the soliton inferred from a Fourier analysis of
the amplitude of the field at the center of the well.
However, the fluctuation of the soliton in the well
is anharmonic.

\begin{figure}[tb]
\epsffile{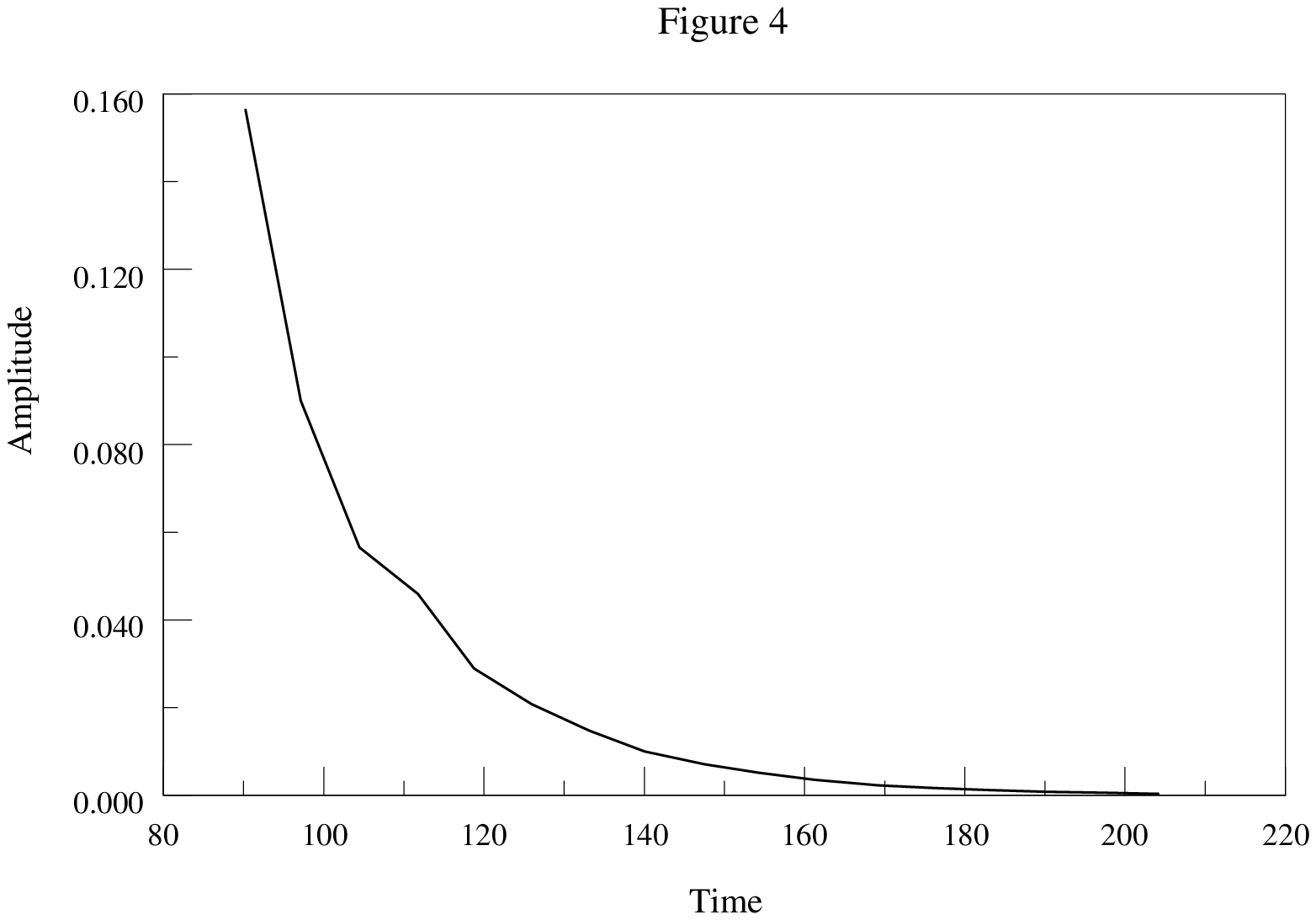}
\label{fig4}
\end{figure}
Another way to observe the decay of a trapped state to a bound state
consists in 
adding a dissipative force of the form $\gamma \frac{\dd\phi}{\dd t}$.
This force cannot be derived from a Hamiltonian, but, it can arise
from the interaction to a bath.
Inserting this term in the soliton equation of motion yields the 
results depicted in figure 4, where we took 
the same set of  parameters as those of
the radiation run of figure 3, but with a friction coefficient $\gamma=.1$.

Attenuation is the dominant effect in this case. The soliton loses
its energy by dissipation instead of radiating it.
Other choices of parameters may lead to a mixture of both processes.
It is, however, evident that
trapped states will eventually become bound states.

Hence, there should exist static bound state solutions of
the soliton in the well. We found those solutions, by integrating the
static equations of motion starting from the center of the 
well. There appears to be only a single bound state for
each choice of well depth and width.
Two bound state solutions are depicted in figure 5. The soliton
is markedly modified by the potential. The total
energy of the soliton may even become negative, as for the
soliton depicted with the dashed curve in the figure.
For this case, the binding energy exceeds the free soliton mass.

\begin{figure}[tb]
\epsffile{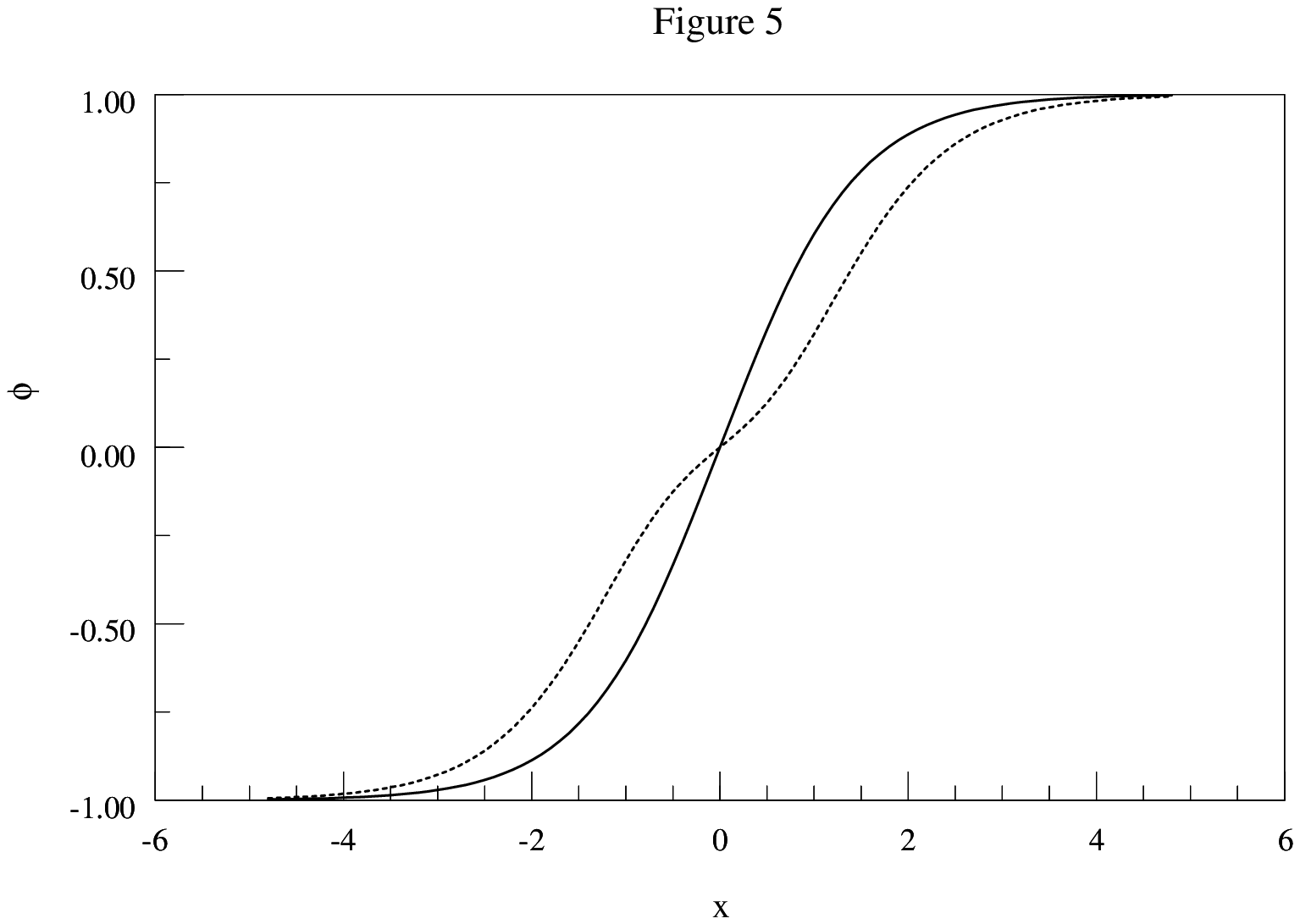}
\label{fig5}
\end{figure}

We have also found static bound state solutions for a soliton located
off-center from the well. Those solutions exist for the soliton located
anywhere on the x-axis.

Several questions arise from the present work and they will be addressed
in future works. One of the intriguing conclusions we can draw is that simple
classical extended objects may have unsuspected behavior, like
trapping, reflection from an attractive potential,
chaotic behavior, etc. 
Turning the process backwards: an extended object,
may it be a soliton or a classical assembly of bound particles, in a trapped
state, can suddenly be freed from it provided some random
interaction causes the reversal of the process of trapping,
a process reminiscent of the decay of metastable states in quantum
mechanics.
\newpage
{\bf Acknowledgements}

This work was supported in part by the Department of
Energy under grant DE-FG03-93ER40773 and by the National Science Foundation
under grant PHY-9413872.

\newpage
{\bf Figure Captions}
\begin{enumerate}
\item[{\bf Fig. 1}:]  Final velocity of the two-particle system 
as a function of the initial velocity
for the parameters $k=1,~\alpha=1,~n=2,~A=2,~\beta=1$ with
a velocity grid dv=.001.
\item[{\bf Fig. 2}:]Same as figure 1, but with finer velocity grid
dv = .0002.
\item[{\bf Fig. 3}:]Amplitude of the oscillation of the soliton in a trapped
state as a function of time. Soliton parameters: $m=1,~\lambda=1$, 
impurity parameters: $h=-3,~a=2,~x_c=3$.
\item[{\bf Fig. 4}:]Same as figure 3 but including attenuation. Friction
coefficient $\gamma~=~.1$.
\item[{\bf Fig. 5}:]Bound state soliton solution 
in a well with parameters h=-.05
and a=.12(solid line), and h=-5, w=1.2(dashed line).
\end{enumerate}
\end{document}